\documentstyle[preprint,aps,amstex]{revtex}
\tightenlines
\begin{document}
\draft
\title{Cluster variation method and disorder varieties of 
two--dimensional Ising--like models}
\author{Alessandro Pelizzola}
\address{Istituto Nazionale per la Fisica della Materia and 
Dipartimento di Fisica del Politecnico di Torino,
c. Duca degli Abruzzi 24, 10129 Torino, Italy}
\date{\today}
\maketitle
\begin{abstract}
I show that the cluster variation method, long used as a powerful
hierarchy of approximations for discrete (Ising-like) two-dimensional
lattice models, 
yields exact results on the disorder varieties which appear when
competitive interactions are put into these
models. I consider, as an example, the plaquette approximation of the
cluster variation method for the square lattice Ising model with
nearest-neighbor, next-nearest-neighbor and plaquette interactions,
and, after rederiving known results, report simple closed-form
expressions for the pair and plaquette correlation functions.  
\end{abstract}


\section{Introduction}

The cluster variation method (CVM) is a powerful hierarchy of
approximations for lattice models of equilibrium statistical mechanics
which has been invented by Kikuchi \cite{Kik1} and more recently
rewritten by An \cite{An} and Morita \cite{Morita}. It is particularly
well suited to analyse complex phase diagrams of discrete classical
models \cite{ptps}, but in some simple cases it is also known to give
exact results. Since the approximations involved amount to neglecting
correlations except for a finite range, exact results are obtained
whenever correlations have a particularly simple structure, as in
tree-like lattices \cite{trees1,trees2,trees3,trees4} or
one-dimensional strips \cite{lat99}.

The purpose of the present paper is to study the behaviour of the CVM
in another situation in which correlations are particularly simple,
namely in the case of disorder varieties of two-dimensional Ising-like
models with 
competitive interactions. Disorder varieties are known since the
papers by Stephenson \cite{Steph1,Steph2} and have subsequently been
studied by many authors
\cite{Enting,PeschEmer,PeschRys,Baxter,Rujan,JaekMaill,Georges,Meyer}.
On a disorder variety (which is a suitable subspace in the whole
parameter space of a model) the correlation functions factorize in a simple
way, which leads to an effective dimensional reduction of the
model, so one could expect that the CVM might be particularly accurate
or even exact in such a case. This is indeed the case and I shall
show, giving both general arguments and a detailed 
analysis of a particular model, that the CVM is exact on
disorder varieties. 

The plan of the paper is as follows: in Sec.\ II I shall introduce
disorder varieties and briefly recall some of the results which have
been obtained in the past years; Sec.\ III will be devoted to the
definition and explanation of the CVM; in Sec.\ IV the exactness of
the CVM on disorder varieties will be shown and finally, conclusions
will be drawn in Sec.\ V.

\section{Disorder varieties}

A disorder variety is a subspace of the parameter space of a model
with competitive interactions, lying in the disordered phase, where
the correlations have a particularly simple form and the model can
then be integrated exactly. The first example of such a variety has
been found by Stephenson \cite{Steph1} in the anisotropic
antiferromagnetic Ising model on the triangular lattice. The
hamiltonian of the model can be written in the form
\begin{equation}
H = - \sum_{\langle i j \rangle} J_{ij} \sigma_i \sigma_j,
\end{equation}
where $\sigma_i = \pm 1$ is the spin variable at site $i$, the sum is over
all nearest-neighbour (NN) pairs and $J_{ij}$ depends only on the
direction of the link between sites $i$ and $j$. The values of
$J_{ij}$ along the three lattice directions will be denoted by $J_1$,
$J_2$ and $J_3$. In the antiferromagnetic model we have $J_l <
0$ for $l = 1, 2, 3$. Stephenson showed that when the condition 
\begin{equation}
{\rm tanh} K_3 + {\rm tanh} K_1 {\rm tanh} K_2 = 0, \qquad K_l =
J_l/k_{\rm B}T
\end{equation}
(or one which is obtained from it by a cyclic permutation of the
indices) holds, then the pair correlation along a lattice direction
has a simple exponential form, as for the one-dimensional model. If
$\sigma_i$ and $\sigma_j$ are two spin variables separated by a distance $k$ on
a linear chain of the lattice in the $l$th direction, their
correlation $\langle \sigma_i \sigma_j \rangle$ is the $k$th power of the NN
correlation along the same direction. In particular, assuming $J_1 <
J_2 < J_3 < 0$, one has $\langle \sigma_i \sigma_j \rangle = [{\rm
tanh}(K_1)]^k$ in direction 1, $\langle \sigma_i \sigma_j \rangle = [{\rm
tanh}(K_2)]^k$ in direction 2 and $\langle \sigma_i \sigma_j \rangle = [{\rm
tanh}(-K_3)]^k$ in direction 3. Stephenson also showed that the
disorder variety separates a portion of the disordered phase in
which the pair correlation has an oscillating behavior from one in
which it decreases monotonically. Similar results have been obtained
by the same author \cite{Steph2} for the union jack lattice and for certain
one-dimensional lattices. 

Later, Enting \cite{Enting} showed that the interaction round a face
(IRF) model on the square lattice (and in particular the Ising model 
with NN, next-nearest-neighbour (NNN) and plaquette
interactions) has a disorder variety which can be mapped onto an
exactly solvable crystal growth model. Peschel and Emery
\cite{PeschEmer} rederived Stephenson's results for the correlations
on the disorder variety of the triangular Ising model by means of a 
one-dimensional kinetic model and applied this technique also to the
ANNNI model. Peschel and Rys \cite{PeschRys} solved the eight vertex
model on one of its disorder varieties. 

Baxter \cite{Baxter} analysed the disorder varieties of the
IRF model on the square lattice. He
showed that the eigenvector of the (diagonal to diagonal) transfer
matrix corresponding to the largest eigenvalue can be written in a
simple form as the product of a sequence of two-site (NN) factors. 

Ruj\`an \cite{Rujan} studied the relations between different techniques and
considered several models (vertex models, staggered IRF model,
$q$-state Potts models, random bond models). 

Jaekel and Maillard \cite{JaekMaill} found a local criterion which
characterizes disorder varieties for any dimensionality and explains
the effective dimensional reduction occurring in the model: the
Boltzmann weight of an elementary cell of the lattice, summed over
some (suitably chosen) spins (or whatever degrees of freedom), is
independent of the remaining spins. Georges and coworkers
\cite{Georges} used this local criterion to calculate correlation
functions on the disorder varieties of three-dimensional Ising
models. 

To conclude this (certainly not exhaustive) brief survey of the
existing literature, we mention that recently, Meyer and coworkers
\cite{Meyer} studied the disorder varieties of the eight vertex model
in the framework of a random matrix theory approach to the transfer
matrix.

\section{The cluster variation method}

The cluster variation method (CVM) is a hierarchy of approximation
techniques for discrete classical lattice models, which has been
invented by Kikuchi \cite{Kik1}. In its modern formulation
\cite{An,Morita} the CVM is based on the truncation of the cumulant
expansion of the variational principle of
equilibrium statistical mechanics, which says that the free energy
${\cal F}$ of a model defined on the lattice $\Lambda$ is given by
\begin{equation}
{\cal F} = {\rm min} \ {\cal F}[\rho_\Lambda] = {\rm min} \ 
{\rm Tr}(\rho_\Lambda H + \rho_\Lambda \ln \rho_\Lambda),
\end{equation}
where $H$ is the hamiltonian of the model, $k_{\rm B}T = 1$ for simplicity,
and the minimization must be performed with respect to a density
matrix obeying the normalization constraint ${\rm Tr}(\rho_\Lambda) = 1$. 

If the model under consideration has only short range interactions and the
maximal clusters are sufficiently large the hamiltonian can be
decomposed into a sum of cluster contributions $H_\alpha$ and the approximate
variational free energy takes the form
\begin{equation}
F[\{\rho_\alpha, \alpha \in M\}] 
= \sum_{\alpha \in M} \left[ {\rm Tr}(\rho_\alpha H_\alpha) -
a_\alpha S_\alpha \right],
\label{CVMfree}
\end{equation}
where $\alpha$ is a cluster of sites,
$\rho_\alpha = {\rm Tr}_{\Lambda \setminus \alpha} \rho_\Lambda$
is the cluster density matrix (${\rm Tr}_{\Lambda \setminus \alpha}$
denotes a summation over all degrees of freedom 
except those belonging to the cluster $\alpha$),
$S_\alpha = - {\rm Tr} (\rho_\alpha \ln \rho_\alpha)$
is the cluster entropy and
the coefficients $a_\alpha$ can be easily obtained from the set
of linear equations \cite{An,Morita} 
\begin{equation}
\sum_{\beta \subseteq \alpha \in M} a_\alpha = 1, \qquad \forall
\beta \in M.
\label{CVMcoefs}
\end{equation}
The cluster density matrices must satisfy the following
normalization and compatibility conditions
\begin{equation}
{\rm Tr} \rho_\alpha = 1, \forall \alpha \in M \qquad {\rm and} \qquad
\rho_\alpha = {\rm Tr}_{\beta \setminus \alpha} \rho_\beta, \forall
\alpha \subset \beta \in M.
\label{CVMtraces}
\end{equation}

Notice that (\ref{CVMfree}) would still be exact if the density matrix
$\rho_\Lambda$ of 
the whole lattice could be written exactly as a product of cluster
density matrices in the form
\begin{equation}
\rho_\Lambda = \prod_{\alpha \in M} (\rho_\alpha)^{a_\alpha}.
\label{rhoprod}
\end{equation}

\section{Exactness of the cluster variation method on disorder varieties} 

There are two properties of the disorder varieties which
suggest, at least for two-dimensional models, that the CVM might be
exact on them. One is the one-dimensional-like character of the pair
correlations. In fact, it is known that for a one-dimensional model
with NN interactions, the pair approximation of the CVM (that is, the
approximation in which the maximal clusters are the NN pairs), which
is equivalent to the Bethe-Peierls approximation, is exact
\cite{Kik1,Brascamp}. 

The other property, still valid for two-dimensional models, is related
to a result by Baxter \cite{Baxter}. He showed that the eigenvector
(corresponding to the largest eigenvalue) of the diagonal to diagonal
transfer matrix is simply the product of a sequence of two-site (NN)
factors. Since the density matrix of a diagonal cluster is the square
of this eigenvector also the density matrix has a product
structure. As we have seen in the previous section, when the density
matrix has a suitable product structure the CVM becomes
exact. Therefore one can hope to find a CVM approximation which is
exact on the disorder variety of a given two-dimensional model. In the
square lattice case a good candidate is the plaquette approximation
\cite{Kik1,Sanchez}, which is equivalent to the Kramers-Wannier
approximation \cite{KW}, which in turn has long been known to
correspond to a variational approximation in which the largest
eigenvalue of the transfer matrix is searched using a restricted space
of factorized vectors \cite{BaxterVar}.

In the case of the plaquette approximation for 
a model defined on the square lattice the condition (\ref{rhoprod}),
which implies the exactness of the approximation, becomes
\begin{equation}
\rho_\Lambda = \frac{\prod_{\rm plaq} \rho_{\rm plaq} \prod_{\rm
site} \rho_{\rm site}}{\prod_{\rm pair} \rho_{\rm pair}},
\label{rhoprodSQ}
\end{equation}
where $\rho_\Lambda$ denotes the density matrix of the whole lattice
and the products are to be intended over all plaquettes, pairs and
sites of the lattice. The above equation should however be taken with
some care, since it is known that not all local thermodynamic states
(i.e. density matrices) can be extended to the whole lattice
\cite{Schlijper}. Consider as an example a model of Ising spins
$\sigma_i = \pm 1$ in its disordered phase, which will be studied in
detail below. One
can easily check on small lattices that, using a generic plaquette
density matrix and the pair and site matrices derived from it by
partial traces, (\ref{rhoprodSQ}) leads to a $\rho_\Lambda$ which
is not correctly normalized. In the case of open boundary conditions
(with this choice the sites and the pairs lying at the boundary do not
enter the products in (\ref{rhoprodSQ})) the correct normalization
is achieved only if $d = c^2$, where $c = \langle \sigma_i \sigma_j
\rangle_{\rm NN}$ and $d = \langle \sigma_i \sigma_j
\rangle_{\rm NNN}$ are the NN and NNN correlations, respectively.

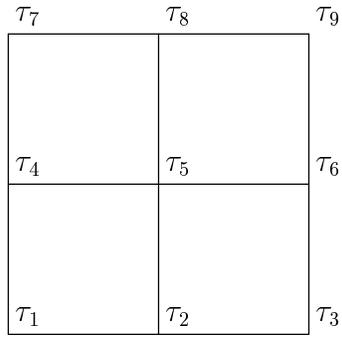
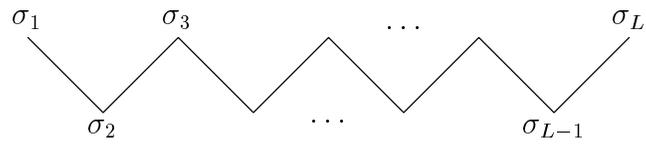
\begin{figure}
\caption{The $3 \times 3$ square and the zig-zag chain}
\label{DiagSqLat}
\end{figure}

When the condition $d = c^2$ holds, the procedure of extending local
density matrices to larger clusters is well defined, in the sense that
by partial traces one can reobtain the local density matrices which were
used to build the larger ones. In addition, one can verify that the
density matrix of any cluster admits a decomposition into a product of
plaquette, pair and site density matrices, with exponents given by the
CVM rules. For instance, with reference to Fig.\ \ref{DiagSqLat}, in
the case of the $3 \times 3$ square we have 
\begin{equation}
\rho_{9}(\tau_1,\ldots \tau_9) = \frac{
\rho_{\rm plaq}(\tau_1,\tau_2,\tau_5,\tau_4)
\rho_{\rm plaq}(\tau_2,\tau_3,\tau_6,\tau_5)
\rho_{\rm plaq}(\tau_4,\tau_5,\tau_8,\tau_7)
\rho_{\rm plaq}(\tau_5,\tau_6,\tau_9,\tau_8)
\rho_{\rm site}(\tau_5)}
{\rho_{\rm pair}(\tau_2,\tau_5) \rho_{\rm pair}(\tau_5,\tau_8)
\rho_{\rm pair}(\tau_4,\tau_5) \rho_{\rm pair}(\tau_5,\tau_6)},
\end{equation}
while for the zig-zag chain 
\begin{equation}
\rho_{\rm chain}(\sigma_1, \sigma_2, \ldots \sigma_L) = 
\frac{\rho_{\rm pair}(\sigma_1,\sigma_2) \rho_{\rm pair}(\sigma_2,\sigma_3) 
\ldots \rho_{\rm pair}(\sigma_{L-1},\sigma_L)}
{\rho_{\rm site}(\sigma_2) \rho_{\rm site}(\sigma_3) 
\ldots \rho_{\rm site}(\sigma_{L-1})}.
\label{zigzagrhoprod}
\end{equation}
As a consequence, also the pair correlation function has a very simple
product form, that is (labeling the spin variables by the site
coordinates) 
\begin{equation}
g(x,y) = \langle \sigma(x_0,y_0) \sigma(x_0+x,y_0+y) \rangle = c^{|x|+|y|}.
\label{corfun}
\end{equation}

The result (\ref{zigzagrhoprod}) is equivalent to the result by
Baxter \cite{Baxter} 
that, on disorder varieties, the eigenvector of the diagonal to
diagonal transfer matrix, 
corresponding to the largest eigenvalue, can be written as a product
of NN pair terms. $\rho_{\rm chain}$ is just the square of this
eigenvector, and the site factors which appear in the denominator can
be easily associated, in a symmetric way, to the adjacent pairs. 
This shows (although this is not a rigorous proof) that when
the plaquette approximation is exact for a model of Ising spins on the
square lattice, then the model is at a point of the disorder variety
in its parameter space.

Let us finally study in detail the square lattice Ising model with NN,
NNN and plaquette interactions. 
The hamiltonian of the model can be written in the form 
\begin{equation}
H = - J_1 \sum_{\langle i j \rangle} \sigma_i \sigma_j 
- J_2 \sum_{\langle \langle i j \rangle \rangle} \sigma_i \sigma_j
- J_4 \sum_{^j_k\square^i_l} \sigma_i \sigma_j \sigma_k \sigma_l,
\end{equation}
where $J_1$, $J_2$ and $J_4$ are the NN, NNN and plaquette couplings,
respectively. This is a special case of the models
studied in \cite{Enting,Baxter,Rujan,Meyer}. We shall first use the
plaquette approximation of the CVM.  Notice that this approximation
has already been applied to the same model in
\cite{Sanchez,Moran,Buzano,Cirillo}. In particular, Sanchez
\cite{Sanchez} reported closed-form expressions for the equilibrium
density matrices and the momentum space pair correlation function in
the disordered phase. Mor\'an-L\'opez and coworkers \cite{Moran}
observed qualitatively the existence of a disorder locus in the phase
diagram. Cirillo and coworkers \cite{Cirillo} calculated again
the momentum space pair correlation function, and on this basis they
determined the location of the disorder line. Their pair correlation
function coincides with that by Sanchez except for a misprint
\cite{Cirillo-priv}, 
and they obtained a disorder line which is only very
close to the exact one, instead of coincident as it should be on the
basis of the results of the present paper, because of an additional
approximation. 

As a first step one can, at least at the numerical level, verify that the
approximation is exact on the disorder variety using only the CVM. A
simple way is to consider a hierarchy of approximations like the
so-called C-series \cite{KikBrush}, in which the maximal cluster is a
rectangle made of $2 \times L$ sites, with $L \ge 2$ (the plaquette
approximation is the first element of the C-series). It is found,
with extremely high precision, that a sequence of approximations
in this series gives identical results on the disorder variety of the
model. Inspection of the pair correlations shows that (\ref{corfun})
is also satisfied.

On the other hand, using published results \cite{Sanchez} it requires
only a long but straightforward calculation to check that on the
(known \cite{Enting,PeschRys,Meyer}) disorder variety of the model one
obtains the exact free energy. Looking at the pair correlations one
also sees that the condition $d = c^2$ (see (\ref{corfun})) is
satisfied on the variety of equation
\begin{equation}
\cosh (2 J_1) = \frac{\exp(2J_4)\cosh(4J_2)+\exp(-2J_2)}
{\exp(2J_2)+\exp(2J_4)},
\end{equation}
which is precisely the disorder variety of the model
\cite{Enting,Meyer}. The free energy per site can be written as 
\begin{equation}
f = - \ln \left[ \exp(-J_4)+\exp(J_4 - 2J_2) \right],
\end{equation}
and again coincides with the exact one, while the NN correlation is 
\begin{equation}
c = \frac{\exp(-4J_2) - \cosh(2 J_1)}{\sinh(2J_1)},
\end{equation}
the NNN correlation $d = c^2$ and the plaquette correlation
\begin{equation}
q = \langle \sigma_i \sigma_j \sigma_k \sigma_l\rangle = 
\frac{\exp(4J_4)\left[1-\exp(8J_2)\right] +
4\exp(2J_2)\left[\exp(2J_4)-\exp(2J_2)\right]} 
{\exp(4J_4)\left[1-\exp(8J_2)\right] +
4\exp(2J_2)\left[\exp(2J_4)+\exp(2J_2)\right]}.
\end{equation}

Finally, since all the pair correlations are given simply by 
(\ref{corfun}) we can easily calculate the momentum space correlation
function, or structure factor. We first rewrite (\ref{corfun}) as
$g(x,y) = \exp\left(-\displaystyle\frac{|x|+|y|}{\xi}\right)$, where
$\xi = -(\ln c)^{-1}$. After a Fourier transform one finds $S(p_x,p_y)
= S_1(p_x) S_1(p_y)$, where
\begin{equation}
S_1(p) = \frac{\sinh(1/\xi)}{\cosh(1/\xi) - \cos p}.
\end{equation}
It can be verified that the structure factors calculated by Sanchez
\cite{Sanchez} and (except for the misprint) Cirillo and coworkers
\cite{Cirillo} reduce to the above expression on the disorder line.

\section{Conclusions}

I have shown that the CVM gives exact results on
the disorder varieties of two-dimensional Ising-like models. In
particular, I have considered the Ising model with NN, NNN and
plaquette interactions on the square lattice, in the plaquette
approximation of the CVM. In the disordered phase of the model,
imposing the simple condition that the NNN pair correlation equals the
square of the NN pair correlation, the CVM plaquette approximation
becomes exact and it is shown that this condition holds on the
disorder variety, where the model can be solved in closed form. It is
important to notice that, using the CVM, one can obtain any
correlation function, due to the fact that the procedure of extending
the local thermodynamic state is well-defined just on the disorder
variety.  Similar results can be obtained
on the triangular lattice as well as other two-dimensional lattices.

\end{document}